# ROSF: Leveraging Information Retrieval and Supervised Learning for Recommending Code Snippets

He Jiang*, Liming Nie, Zeyi Sun, Zhilei Ren, Weiqiang Kong, Tao Zhang, and Xiapu Luo

**Abstract**—when implementing unfamiliar programming tasks, developers commonly search code examples and learn usage patterns of APIs from the code examples or reuse them by copy-pasting and modifying. For providing high-quality code examples, previous studies present several methods to recommend code snippets mainly based on information retrieval. In this paper, to provide better recommendation results, we propose ROSF, Recommending cOde Snippets with multi-aspect Features, a novel method combining both information retrieval and supervised learning. In our method, we recommend Top-*K* code snippets for a given free-form query based on two stages, *i.e.*, coarse-grained searching and fine-grained re-ranking. First, we generate a code snippet candidate set by searching a code snippet corpus using an information retrieval method. Second, we predict probability values of the code snippets for different relevance scores in the candidate set by the learned prediction model from a training set, re-rank these candidate code snippets according to the probability values, and recommend the final results to developers. We conduct several experiments to evaluate our method in a large-scale corpus containing 921,713 real-world code snippets. The results show that ROSF is an effective method for code snippets recommendation and outperforms the-state-of-the-art methods by 20% - 41% in Precision and 13% - 33% in NDCG.

**Index Terms**—Code snippets recommendation, information retrieval, supervised learning, topic model, feature.

——————————— ◆ ———————————

## 1 INTRODUCTION

INTERNETWARE is a software paradigm consisting of self-contained, autonomous entities in Internet computing environment [30]. As mentioned in previous work, both desktop software and mobile applications (apps) are possible entities in Internetware systems [30], [22]. In the development process for these software, developers often have to implement unfamiliar programming tasks. They either reuse code examples by copy-pasting and modifying [23], or learn the correct ways to employ an unfamiliar Application Programming Interface (API) relying on code examples [54]. As one of the most common ways for reuse, code reuse can save time and resources and reduce redundancy [32].

A code snippet refers to a piece of code, which can accomplish one or more specific programming tasks [17]. Typically, a programming task, for example "record sound audio", is a short text that describes the requirements on the program to be constructed. To find high-quality code examples for programming tasks, developers may search the publicly available code repositories on the Internet or locally available projects [28]. Some Internet-scale code search engines, such as Open Hub [4], can provide code examples for a given task. However, the dominant measure used by these engines is textual similarity [11]. Previous studies show that these results are usually complicated and not sufficient [17].

In recent years, some researchers propose several methods to recommend code snippets for free-form queries[7], [17], [29]. These methods rank the code snippets in a corpus and return Top-*K* related code snippets to developers. An earlier study [23] shows that the performance of these methods has room for improvement. The possible reasons may include that a signal feature is used for ranking and the weights of features cannot be adjusted automatically. The features employed in these methods contain textual similarity between a query and code snippets [29], code metrics such as the lines of code [36], etc. For achieving better performance, it is necessary to employ multiple features and assign different weights for these features automatically [7], [36]. Supervised learning can handle this scenario above, which is the machine learning task of inferring a model from labeled training set. Using the learned prediction model, one can determine the class labels for unseen instances in a test set for a new query [31], [50], and further recommend relevant code snippets.

In this paper, we propose Recommending cOde

————————————————

- *H. Jiang is with the School of Software, Dalian University of Technology, Dalian, China and the Key Laboratory for Ubiquitous Network and Service Software of Liaoning Province, Dalian, China, is also with the State Key Laboratory of Software Engineering, Wuhan University, Wuhan, China. E-mail: jianghe@dlut.edu.cn.*
- *L. Nie, Z. Sun, Z. Ren, and W. Kong are with the School of Software, Dalian University of Technology, Dalian, China and the Key Laboratory for Ubiquitous Network and Service Software of Liaoning Province, Dalian, China. E-mail: limingnie@mail.dlut.edu.cn; sunzeyidlut@gmail.com; {zren, wqkong}@dlut.edu.cn.*
- *T. Zhang is with School of Computer Science & Technology, Nanjing University of Posts and Telecommunications, Nanjing, China. Email: cstzhang@njupt.edu.cn*
- *X. Luo is with Department of Computing, The Hong Kong Polytechnic University, Hong Kong, China. Email: csxluo@comp.polyu.edu.hk.*

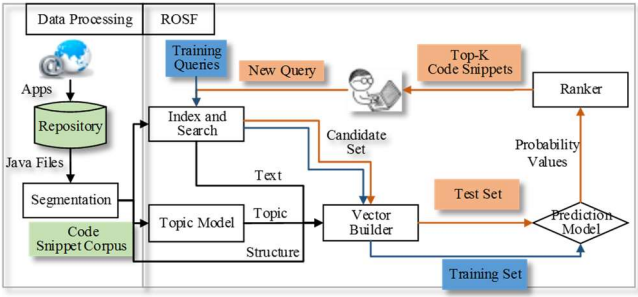

Fig. 1. Code snippets recommendation framework. Two phases: Data Processing and recommendation (ROSF). The blue arrows indicate the process of training. The orange arrows indicate the process of recommending for a new query.

Snippets with multi-aspect Features (ROSF). Our method combines both information retrieval and supervised learning to improve the performance of code snippets recommendation. In this method, we explore two stages, *i.e.*, coarse-grained searching and fine-grained re-ranking, to recommend code snippets. First, for a new free-form query, a candidate set is generated by using an information retrieval method. This stage tries to collect as many as possible relevant code snippets in the candidate set. Then, we re-rank the candidate set by a learned prediction model. This stage tries to re-rank the relevant code snippets to the top of the results by the supervised learning method. Here, the problem of code snippets recommendation is viewed as a multi-class classification. The prediction model is learned on the instances of a training set by utilizing multinomial logistic regression [6]. Each instance refers to a candidate code snippet, and is represented with a vector containing multi-feature values and a label. These features reflect three aspects of code snippets, *i.e.*, text, topic, and structure. The label reflects the relevance score between a code snippet and the query.

To evaluate the effectiveness of our method, we propose three Research Questions (RQs) and conduct several experiments to answer them. As code reuse is relatively common in mobile apps [32], our experiments are based on a code snippet corpus with more than 920,000 real-world code snippets from 1,538 open source app projects on the Android platform. Moreover, we employ 35 queries and their candidate set to create the training and the test set. Among them, 20 queries are randomly selected as testing queries, the others are treated as training queries. We label each candidate set for each query. Totally, 3,500 instances related to 35 queries are labeled by assessors. The results of experiments show that (1) ROSF can optimize the ranking of the candidate set to achieve better results. (2) ROSF is a better method for code snippets recommendation than comparative methods, which outperforms Portfolio [29] and VF [17] by 20% - 41% in Precision@10 and 13% - 33% in Normalized Discounted Cumulative Gain (NDCG)@ 10.

This paper makes the following contributions:

1. We propose ROSF, a new hybrid code recommendation method based on information retrieval and supervised learning. Our method considers full advantage of text, topic, and structure aspects of code snippets.

2. We evaluate the performance of ROSF against several comparative methods in terms of Precision and NDCG.

3. We explore the impact of features on the performance of ROSF, and present the influential features for supervised learning method to recommend code snippets.

4. We construct a code snippet corpus segmented from open source app projects, and label a set with 3,500 instances for 35 real world free-form queries.

Next section outlines the architecture of our method and a prototype. Section 3 elaborates on the data processing. The steps of training and recommendation are proposed in Section 4. Section 5 provides details about the experimental design. Experimental results are presented in Section 6. Section 7 states the threats to validity. The related works are shown in Section 8. In Section 9, we conclude this paper and introduce the future work.

## 2 ARCHITECTURE AND PROTOTYPE

This section first introduces the overall architecture of our framework, and then shows the whole process using a prototype.

### 2.1 Architecture

Fig. 1 shows two phases of our framework: *Data Processing* and *recommendation (i.e., ROSF)*.

In the *Data Processing* phrase, we input the open source app projects collected from the website F-droid [1], and output a *code snippet corpus.* First, we extract the Java files from app projects and store them in a *Repository*. Then, *Segmentation* parses each of the Java files to generate the *code snippet corpus*. Each method in the Java file is segmented to a code snippet [17]. In the phase of *recommendation*, the input contains the new query and the collected *code snippet corpus*. The output is a list with K ranked code snippets. As it is a time-consuming work to label relevance scores for all code snippets in the code snippet corpus, following Niu et al. [36], we explore two stages to recommend code snippets: coarse-grained searching and fine-grained re-ranking [37].

Specifically, for a new query, we first identify a candidate set that contains *N* code snippets using an information retrieval method (*e.g.*, BM25). Then, the candidate code snippets are represented as instances by *Vector builder* as a test set. Finally, by the learned prediction model from a training set, we predict the probability values belonging to different relevance scores for each instance in the test set. According to the probability values, we identify the predicted relevance

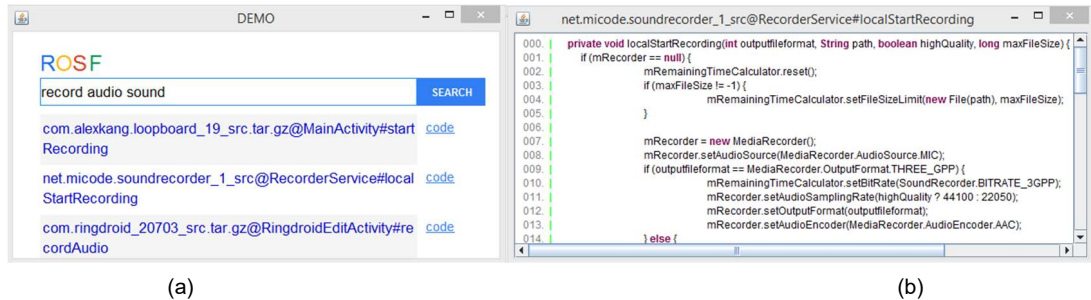

Fig. 2. Prototype of ROSF. Two interfaces are contained: the main user interface (a), and the code snippet display interface (b).

TABLE 1
THE FEATURES AND THEIR CATEGORIES

| ID | Categories | Features |
|---|---|---|
| f1 | | The textual similarity between a query and the content of a candidate code snippet (*i.e.*, code text and comments). |
| f2 | | The textual similarity between a query and the full title of a candidate code snippet (*i.e.*, package name of app, class name, and code snippet name). |
| f3 | | The textual similarity between a query and the simple title (only contains code snippet name) of a candidate code snippet. |
| f4 | Text | The textual similarity between a query and the sibling method names contained in a same Java file with a candidate code snippet. |
| f5 | | The textual similarity between a query and the import statements of Android library in the Java file that contains a candidate code snippet. |
| f6 | | The textual similarity between a query and the import statements of the Java standard library in the Java file that contains a candidate code snippet. |
| f7 | | The textual similarity between a query and the import statements of other libraries in the Java file that contains a candidate code snippet. |
| f8 | Topic | The topic similarity between a query and the content of a candidate code snippet. |
| f9 | Structure | The number of lines in a candidate code snippet. |

score, and recommend a list with *K* (*K<N*) code snippets by the module of *Ranker*.

For learning a prediction model from the training set, we first prepare several training queries and their candidate sets following the steps above. Then, we construct the instances for each training query to generate the training set. One difference from the test set is that the relevance score for each instance is labelled manually by assessors. Finally, we employ multinomial logistic regression [6] to train and generate the prediction model. Recently, there are more applications of IR in software engineering (e.g., for regression test prioritization [41]), it seems that the proposed technique may also be possible to applied to those new areas. It would be interesting to discuss such potentials.

### 2.2 Prototype

To display our method in a more visually appealing manner, we implement a prototype. As shown in Fig. 2, there are two windows: *the main user interface (a)* and *the code snippet display window (b)*.

When a developer enters a free-form query in the search bar of the *main user interface*, for example, "*record audio sound*", ROSF returns a list with 10 code snippets below the search bar. Suppose the code snippet on the second position is selected, by clicking the "*code*" button on the right, the text content of this code snippet will be displayed in the *code snippet display window*. Then, the developer can check the details about this code snippet and get inspiration from it.

The purpose of our work is saving developers' time for searching more relevant code snippets according to their programming tasks. After achieving these code snippets, developers still need to manually modify these code snippets and further test. In other words, we only provide relevant code snippets to developers without considering integration with a code context.

## 3 DATA PROCESSING

In this section, we show the data and the process to acquire them. Then, we present some techniques used in the process, such as BM25, topic model. As mentioned by Mei et al. [30], in the Internet computing environment, mobile apps are possible entities in Internetware systems [22], [53]. Code reuse is relatively common in mobile apps [32], and mobile apps are becoming more popular, we use the data on the Android platform to evaluate our method.

### 3.1 Open Source App Projects

The code snippets recommended in our experiments come from open source app projects. These projects are collected from F-droid [1]. F-droid is a Website with free and open source apps on the Android platform. For each one of the app projects, there are several versions. We only focus on the latest version in our experiments. Until July 2015, we collected all 1,538 Android app projects from F-droid. Following the steps of [35], we extract 921,713 code snippets from these app projects.

## 3.2 Feature Extraction

For each Java file in the app project, we parse it using the module *Segmentation*, and extract three aspects of code snippets: text, topic, and structure. These aspects are used to calculate three types of feature values of instances to train the prediction model, some of these features are used in earlier studies [17], [34], [54]. Table 1 shows some details about these features. Specifically, the text and topic aspects are used to calculate the query-dependent textual similarity features and topic similarity features [21], respectively. A code snippet is considered as a textual document describing one or more technical issues/topics. Each of them is represented by certain words or terms. Intuitively, the terms are textual, visible, while the topics are semantic, latent. They can complement each other to achieve better performance for text matching tasks [24]. Moreover, the structure aspect is employed as query-independent feature. Fig. 3 shows a Java file that implements a programming task "*record sound audio*". Next, we show how to segment Java files and collect these aspects using this example.

**(1) Text**

*Content:* In Fig. 3, there are four methods in the Java file "RecorderService". The tool Eclipse Abstract Syntax Tree (AST) is exploited to parse each Java file [17]. Totally, we can segment four code snippets from this Java file. Each method in the Java file can be segmented as a code snippet, which contains the code text and comments. Following the process, we can generate a code snippet corpus. As shown in Table 2, the first row provides the statistical information about the number of code snippets in our code snippet corpus. Here, "Max/Min" refers to the maximum and minimum number of code snippets that an app project contains. The max value reaches 22,234, which means that project is quite large. The feature f1 in Table 1 is the textual similarity between a query and the content of a candidate code snippet.

*The title of code snippet:* We name each code snippet using a fixed format. In this example, we set the title of the method "localStartRecording" as "net.micode. soundrecorder_1_src@RecorderService#localStartReco rding.txt", where "net.micode.soundrecorder_1_src" refers to the package name of the app, and "RecorderService" is the name of the Java file. We employ two types of titles to calculate the features f2 and f3 in Table 1. The differences are the usages of package names and Java file names.

*The names of sibling methods:* Except for the title of the candidate method, other sibling methods in this Java file also show helpful information. The feature f4 provides the textual similarity between the queries and the names of these sibling methods.

*The import statements:* Beside above information, we also take into account the import statements of Java files [7]. The import statements in Fig. 3 direct the Java compiler to include the android and Java APIs in the compilation. We divide these import statements into three categories: Android, Java, and other libraries. The features f5 - f7 are the textual similarity between a query and the import statements.

**(2) Topic**

In Table 1, f8 is the topic feature. In our experiments, except for the textual similarity above, we also calculate the topic similarity between queries and code snippets using Latent Dirichlet Allocation (LDA) [8]. Given a collection of code snippets and queries, we first generate a term-by-document matrix M. A generic entry $\omega_{ij}$ of this matrix denotes a measure of the weight (*i.e.*, relevance) of the $i^{th}$ term in the $j^{th}$ document. Then, LDA takes the term-by-document matrix as an input to identify the latent variables (topics) hidden in the data and generates as output a matrix $\theta$, denoted as topic-by-document matrix. A generic entry $\theta_{ij}$ of this matrix denotes the probability of the $j^{th}$ document to belong to the $i^{th}$ topic. The number of topics is usually much smaller than the number of terms. Finally, the topic similarity between queries and code snippets can be calculated based on their topic-by-document matrix [38]. In this process, we use the collapsed Gibbs sampling on Mallet [3]. Moreover, following [38], we set the topic number as 100, and the number of iterations as 100.

**(3) Structure**

*The number of lines:* In our preliminary experiments, we observe that the number of lines in code snippets is also an important type of information. Too much or too little line numbers will unsuitable to implement programming tasks. The second row of Table 2 provides the statistical information about the line numbers of code snippets in our code snippet corpus. Because the number of lines in code snippets is query-independent, we use it as feature f9 directly.

## 3.3 Index and Search

By the module *Index and Search* on Lucene, we index the text aspects (*i.e.*, the content, the titles, the sibling method names, and the import statements) of code snippets. Meanwhile, the text features f1 - f7 can be calculated when the query is entered. Lucene is a free and open-source information retrieval engine [2]. We employ Lucene as it can provide high quality and fast services for indexing and searching. Moreover, in our work, we also use Lucene to generate the candidate set for a given query.

TABLE 2
THE STATISTICAL INFORMATION FOR CODES AND LINES

|       | Total      | Max    | Min | Mean | StdDev |
|-------|------------|--------|-----|------|--------|
| Codes | 921,713    | 22,234 | 1   | 600  | 1623   |
| Lines | 11,445,768 | 2,222  | 1   | 12   | 22     |

Before indexing the text aspects of a code snippet, we need to preprocess each of them. The text preprocessing is important in the text mining community [47], [49], [51], which contains tokenization, stop words removal, and stemming. Before tokenization, we split the identifiers to terms by Camel-case. For example, the identifier "MediaRecorder" can be split into "Media" and "Recorder" [29]. After above steps, the text aspects the code snippet are now represented as several bags of terms. Then, we store them on Lucene to generate a document. In other words, a document corresponds to a code snippet and consists of a number of fields. Each field stores the content of each processed text aspects.

### 3.4 BM25

On Lucene, we generate the candidate set and calculate the features against a query with the BM25 textual similarity by Lucene automatically. BM25 is a bag-of-words ranking function implemented in Okapi system [40], which has provided very effective retrieval performance in previous TREC experiments [52]. As the queries with several keywords are often short, BM25 can facilitate the retrieval of documents relevant to a short query [13].

Given a query q with terms $t_1, t_2, ..., t_n$, the BM25 similarity between a document $D$ and the query $q$ is [25], [44]:

$$sim(D,q) = \sum_{t \in q \cap D} IDF(t) \cdot \frac{tf(t,D)(k_1+1)}{tf(t,D)+k_1\left(1-b+b\frac{|D|}{avgdl}\right)} \quad (1)$$

where, $tf(t,D)$ is the term frequency of $t$ in the document D, $|D|$ is the length of document D, and $avgdl$ is the average of document lengths in the whole corpus. The parameters $k_1$ and $b$ control the scale of term frequency and document length, respectively. In our experiments, the values of $k1$, $b$ are 1.2 and 0.75, respectively, which are the recommended values in [40].

The Inverse Document Frequency (IDF) of term $t$ in the whole corpus is calculated as:

$$IDF(t) = log\frac{1+(N-n(t)+0.5)}{n(t)+0.5} \quad (2)$$

where, $N$ is the total number of documents, $n(t)$ is the number of documents that contain the term $t$. 0.5 is a smoothing constant to deal with the situation that $n(t)$ is set to 0.

## 4 TRAINING AND RECOMMENDATION

This section details our method. For a given query from a developer, ROSF is responsible for recommending a list of potential relevant code snippets. There are two stages in this process: coarse-grained searching and fine-grained re-ranking. Specifically, the first stage is achieving Top-*N* code snippets as the candidate set. The second stage is optimizing the ranking of this set using a learned prediction model and recommending Top-*K* code snippets to the developers.

### 4.1 Achieving the Candidate Set

To collect Top-*N* code snippets as the candidate set, we score each snippet in the corpus with a BM25 similarity for the given query by using formula (1) on Lucene.

After scoring, the snippets are ranked based on their values. Snippets with high scores are ranked in the top of the final result, which means these snippets are more relevant to the query. Finally, we can get the candidate set with Top-*N* code snippets. It should be noted that in the comparative method BM25, the Top-*K* code snippets out of Top-*N* code snippets are returned as the final result.

In the process, we employ two strategies to filter the results. The first one, following [7], is to remove the code snippets with less than five lines in the results of our method. Another is the removal of potentially duplicate code snippets in the results. Two code snippets are considered potentially duplicated if they have a same method name and a BM25 similarity.

### 4.2 Vector Builder

*Vector Builder* module is employed to construct the instances of the training set or the test set in our experiments. Each instance corresponds to a candidate code snippet for a query. It is represented as a vector with the form < *q*, *F_c*, *L*>, where *q* refers to the given query; $F_c$ contains different feature values of a candidate code snippet *c*; *L* refers to the relevance score between the query and the candidate code snippet. In a real scenario, we only need to evaluate the relevance score for each instance in the training set manually. The labels for the instances in the test set are predicted by the learned prediction model. However, in our experiments, for generating the golden set, we evaluate all instances in the candidate set for 35 queries. Totally, we label 3,500 instances.

### 4.3 Prediction model and Ranker

In this subsection, we show the re-ranking process using the module *Prediction model and Ranker* in Fig. 1. In our work, we view the problem of code snippets recommendation as a multi-class classification. A candidate code snippet may be labeled with four possible scores. Meanwhile, multinomial logistic regression is a classification method that generalizes logistic regression to multiclass problems [6]. Therefore, multinomial logistic regression is a suitable analytic approach to our problem.

Given a set of feature values of an instance, this classification method constructs a linear predictor function to predict the probabilities of several possible labels for this instance. The linear predictor function is represented with a linear combination of the features and the weights of the features [6].

$$score\ (F_c, L) = \beta_L \cdot F_c \qquad (3)$$

where, $F_c$ is the feature vector of instance $c$, $\beta_L$ is a vector of weights (or regression coefficients) corresponding to relevance score $L$, and $score\ (F_c, L)$ is the probability belonging to relevance score L for the instance c. The best values of the weights for our problem are determined from the training set by using stochastic gradient descent algorithm according to the relevance scores labeled manually.

For each instance of the candidate set for a new query, we employ the learned linear predictor function to predict the probabilities of four possible relevance scores. In other words, each instance has four probability values corresponding to four relevance scores. Then, the relevance score with the maximum probability value is selected as the predicted relevance score for the instance. Among the candidate set, we first sort the subset containing the code snippets with predicted *score 4* according to the predicted probability values in descending order. Then, we select Top-*K* code snippets as the final results. If the size of this subset is less than *K*, we consider the subset with *score 3*, until we collect *K* code snippets.

*For example:* In Table 3, there are five instances, *i.e.*, *a, b, c, d, e*, in a candidate set for a query. Assume we need to recommend Top-3 code snippets. Using the learned predictor model, first, we can generate the probability values of four relevance scores for each instance. Then, we identify the predicted score for each instance by the probability values. For the instance *a*, the maximum value is 0.9 which belongs to the score 3. We set the predicted score as 3. Finally, we can achieve four lists corresponding to four scores. We start to select from the list with score 4, until we collect 3 instances. Finally, the result is: *b, c*, and *a*.

TABLE 3
AN EXAMPLE FOR RE-RANKING THE CANDIDATE SET

| Instances | Relevance Scores | | | | Predicted Score |
| | 1 | 2 | 3 | 4 | |
|---|---|---|---|---|---|
| a | 0.1 | 0.0 | **0.9** | 0.0 | 3 |
| b | 0.0 | 0.2 | 0.1 | **0.7** | 4 |
| c | 0.4 | 0.1 | 0.0 | **0.5** | 4 |
| d | 0.0 | 0.0 | **0.6** | 0.4 | 3 |
| e | **0.8** | 0.0 | 0.1 | 0.1 | 1 |

*Code snippet with predicted score 4: b (0.7), c (0.5);*
*Code snippet with predicted score 3: a (0.9), d (0.6);*
*Code snippet with predicted score 2: null;*
*Code snippet with predicted score 1: e (0.8);*
*Top-3 recommendation results: b, c, and a.*

## 5 EXPERIMENTAL DESIGN

In this section, we evaluate the effectiveness of our proposed method by three research questions based on a large-scale real-world data set. Our experiments are conducted on a 3.60 GHz CPU (Intel i5) PC running windows 8.1 OS with 8G memory. We implement our method using Java in Eclipse. All data used in our experiments can be found on our website for comparison [5].

### 5.1 Research Questions

We explore the following Research Questions (RQs). In the section 6, we conduct several experiments to answer three RQs.

*RQ1: Will the performance of ROSF be affected by the size of the candidate set?*

As mentioned before, in ROSF, we use two stages to recommend code snippets, *i.e.*, coarse-grained searching and fine-grained re-ranking. In coarse-grained searching, we identify a candidate set with N code snippets. Here, we propose RQ1 to explore the impact of the parameter N on the performance of our method.

TABLE 4
QUERIES FOR TEST

| ID | Query | Tag | Viewed times |
|---|---|---|---|
| 1 | Record audio sound | android | 4783 |
| 2 | Get screen dimensions in pixels | android, layout, screen | 720031 |
| 3 | Take a screenshot on Android | android, screenshot | 107071 |
| 4 | Get the memory used | android, memory, memory-management | 217026 |
| 5 | Get the list of activities/applications installed | android | 180820 |
| 6 | Import the system time | android, operating-system | 36113 |
| 7 | Open a URL in Android's web browser | android, url, android-intent, android-browser | 342424 |
| 8 | Use android Timer in Android activity | android, multithreading, timer, scheduled-tasks | 18998 |
| 9 | Capture Image from Camera and Display in Activity | android, image, camera, capture | 157947 |
| 10 | Handle right to left swipe gestures | android, swipe, gesture-recognition | 152674 |
| 11 | Converting pixels to dp | android | 264672 |
| 12 | Draw a line in android | android | 182837 |
| 13 | Get cpu usage | android, cpu-usage | 72209 |
| 14 | Detect network connection status | android, networking, wifi, connectivity | 69553 |
| 15 | Check if an application is installed or not in Android | android, apk | 46174 |
| 16 | Convert an image into Base64 string | android, Base64 | 80049 |
| 17 | Get the web page contents from a WebView | android, android-webview | 52124 |
| 18 | Cancel an executing AsyncTask | android, android-asynctask | 85973 |
| 19 | Detect if a Bluetooth device is connected | android | 39245 |
| 20 | Retrieve incoming call's phone number | android, telephonymanager, phone-state-listener | 50134 |

*RQ2: Will the performance of ROSF be better than comparative methods?*

Different from previous methods mainly using information retrieval method, we are trying to employ both information retrieval and supervised learning to recommend code snippets. In this research question, we want to explore the performance of our method by comparing with several state-of-the-art methods.

*RQ3: How does each of the features affect the performance of our method?*

This research question is proposed to evaluate the impact of each feature on the performance of our method. Before that, we first explore the correlation between these features.

### 5.2 Queries

A query refers to a programming task. In order to simulate the real scenario, following the method in [23], we collect real-world programming tasks from Stack Overflow [4] as queries. We totally employ 35 queries in our experiments. Among them, 20 queries are randomly selected for test, the others are treated as training queries. In other words, the instances of the test set refer to the candidate sets of 20 test queries. Meanwhile, the instances of the training set refer to the candidate sets of 15 training queries. We use the same test queries for all methods in the following experiments.

Table 4 shows some details about the test queries. All 35 queries can be found in our webpage [5]. The column "*Tag*" shows the categories of the queries. Note that these queries share the tag "android", which means that they are related with mobile apps development. The column "*Viewed times*" indicates the number of times a query has been viewed by visitors. These values are all comparatively large, which means that the visitors desire to achieve the solutions of these programming tasks.

For collecting these programming tasks, we first manually rank the posts with the "android" tag on Stack Overflow. Then, we check the posts one by one following some criteria until we collect 35 tasks. Following [29], the criteria are that the tasks should belong to Android app development and be viewed many times. Meanwhile, there are solutions (*i.e.*, the accepted answers) along with these programming tasks in the same webpages. These solutions can assist in evaluating the relevance scores of code snippets. Finally, we extract the titles of these tasks as queries.

### 5.3 Evaluation

For evaluating the relevance score between a code snippet and a query, we conduct the following evaluation process [55]. First, for each test query, we obtain the Top-*K* code snippets from each comparative method. Then, we merge all code snippets into a pool which includes only unique code snippets. For each code snippet in this pool, we recruit two assessors to evaluate the relevance score between a code snippet and the query. As regards the inconsistencies of labeling, we recruit an expert to arbitrate the score. Finally, we employ two metrics to measure the performance of each method.

Two assessors are graduate students from Dalian University of Technology, and the expert is a doctoral student from the same school. Both the two assessors and the expert have at least three years of Android app development experience. Meanwhile, two assessors have at least four years of Java development experience, and the expert has more than nine years of Java development experience. Before the evaluating, we give them a 30-minutes training about labeling.

Following the method in [29], we label the relevance scores with a Four-level Likert scale. Meanwhile, the solutions together with the programming tasks in the Stack Overflow are used to assist labeling. In other words, the assessors can check the descriptions about the programming tasks and their solutions in evaluating. Here, we present the guidelines for labeling as follows: *Score 4*: Highly relevant. The code snippet is perfectly suitable for the programming task. *Score 3*: Mostly relevant. The code snippet or the APIs used in this snippet can be reused for the programming task with some changes. *Score 2*: Mostly irrelevant. The code snippet only contains a few relevant code lines, which is not enough to solve the programming task. *Score 1*: Completely irrelevant. The code snippet cannot solve the programming task. To put it simply, if the relevance score of a code snippet is equal to or greater than 3, *i.e.*, 3 or 4, the code snippet should contain useful code lines or APIs to solve the programming task.

TABLE 5
RECOMMENDATION PERFORMANCE OF ROSF WHEN N EQUALS TO DIFFERENT VALUES

| Top-N | Precision@10 | NDCG@10 |
|---|---|---|
| BM25 | 57.5% | 0.7551 |
| 20 | 60.5% | 0.7777 |
| 30 | 60% | 0.8167 |
| 40 | 61.5% | 0.8140 |
| 50 | 62.5% | 0.7942 |
| 60 | 63% | 0.8248 |
| **70** | **66.5%** | **0.8448** |
| 80 | 64% | 0.8427 |
| 90 | 62% | 0.8081 |
| 100 | 60.5% | 0.8158 |

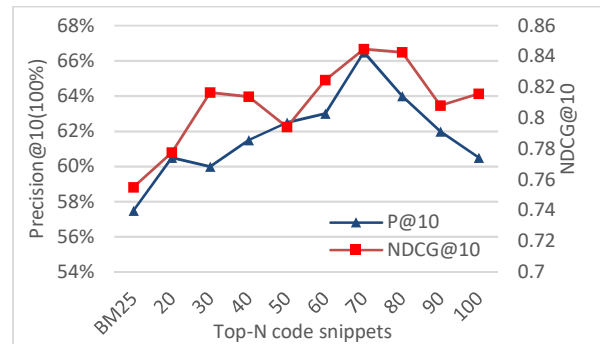

Fig. 4. The trends of Precision@10 and NDCG@10 when the size of the candidate set equals to different values.

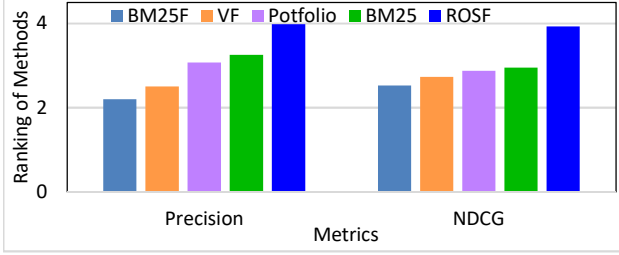

Fig. 5. Comparisons for average ranking of five methods for Friedman's test

### 5.4 Metrics

An ideal recommendation method should hit more of the relevant records and place them at the top of the results. Following the previous methods[17], [29], we evaluate the performance of each method using two metrics, *i.e.*, *Precision@K* and Normalized Discounted Cumulative Gain *(NDCG)@K*. Because we know nothing about the number of relevant code snippets not retrieved for a given query, following the previous method [29], it is impractical to calculate the metric recall.

Specifically, the *Precision@K* is defined as the proportion of the true positives (*i.e.*, the code snippets with score 3 or 4) in Top-*K* recommended results (both true positives and false positives) [29].

The *Precision@K* is calculated as:

$$Precision@K = \frac{|Relevance|}{|Retrieved|} \quad (4)$$

where the numerator |*Relevance*| is the number of relevant code snippets in the result. The denominator |*Retrieved*| is the total number of results recommended by a method, which equals to 10 in our study.

The metric *NDCG* is commonly used in information retrieval to measure the ranking capability of a recommendation method. A method is more useful when there are more relevant results in higher positions in the hit list than irrelevant results. We calculate *NDCG@K* of each method for given queries.

$$NDCG@K = \frac{DCG@K}{IDCG@K} \quad (5)$$

$$DCG@K = R_1 + \sum_{i=2}^{K} \frac{R_i}{log_2 i} \quad (6)$$

where *NDCG@K* is the *DCG@K* normalized by *IDCG@K*. *IDCG@K* is the ideal *DCG@K*, where the results are sorted by relevance scores. $R_1$ is the relevance score at the first position in the list. $R_i$ is the relevance score at the *ith* position.

In the experiments, we observe that the value of NDCG cannot show the real performance for recommendation. For example, there are two results from different methods for a given query. The result A is 4, 1, 1, and 1. The result B is 2, 2, 2, and 2. We recommend 4 code snippets in these results. The values of NDCG for two results are all equal to 1. However, we find that there are no relevant code snippets in the result B (a snippet with score 3 or 4 was considered to be relevant). To solve this problem, we set the score 1 and score 2 to score 0 in the returned results. Then, the NDCG value of the result B equals to 0. The NDCG value of the result A still equals to 1.

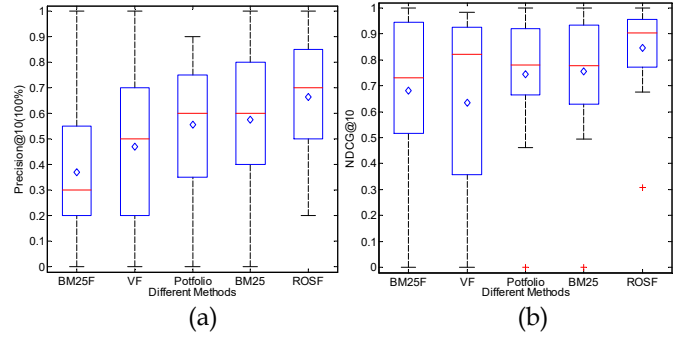

Fig. 6. The statistical results of Precision@10 (a) and NDCG@10 (b) for different methods. The x axes indicate five recommendation methods. The y axes indicate the range for two metrics, respectively. The red line represents the median. And the blue rhombus represents the mean.

## 6 EXPERIMENTAL RESULTS

This section presents the results of several experiments to answer the three RQs proposed in section 5.1.

### 6.1 Answer to RQ1

To answer RQ1, we explore the impact of the parameter N for recommendation, *i.e.*, the size of the candidate set. In this experiment, we use the result generated by BM25 as a baseline, which refers to the result without the re-ranking process. Then, we achieve several results when parameter *N* equals to different values using our method. Finally, the metrics based on these results are calculated and compared.

Table 5 summaries the numerical results of two metrics. In the column Top-*N*, Lines *20* to *100* refer to several results of ROSF when *N* equals to these values. When *N* equals to these values, the performance of ROSF is greater than BM25. We also find that, when *N* equals 70, ROSF obtains the best performance, where the value of Precision@10 is 66.5%, and the value of NDCG@10 is 0.8448. ROSF improves BM25 by up to 16% in Precision, and 12% in NDCG. The results show that ROSF has the ability to improve the positions of the relevant code snippets in the candidate set.

In a more intuitive way, Fig. 4 shows the trends of two metrics. In the figure, we note that the values of two metrics first all rise and then descend. They all reach their peaks when *N* equals to 70. The possible reason may be that more irrelevant code snippets will be added in the candidate set than relevant ones, as the *N* value gradually increases. Finally, we chose 70 as the default value of the parameter *N* in the following

TABLE 6
RESULTS OF WILCOXON'S TESTS BETWEEN ROSF AND OTHER COMPARATIVE METHODS FOR TWO METRICS

| Metrics | R vs B | R vs P | R vs VF | R vs BF |
|---|---|---|---|---|
| Precision@10 | 0.0120 | 0.0059 | 0.0147 | 0.0019 |
| NDCG@10 | 0.0620 | 0.0486 | 0.0057 | 0.0333 |

*R refers to ROSF, B refers to BM25, P refers to Portfolio, and BF refers to BM25F. The first column indicates the two metrics. In other columns, the results of Wilcoxon's tests are reported for each metric. The comparison results consist of the p−value.*

experiments. In this experiment, we only discuss the situations when the values of parameter *N* are less than 100. We still do not know the trends in other cases. This will be a threat to our conclusions.

*Answer to RQ1*: The performance of ROSF varies when the size of the candidate set takes different values. When the *N* is less than 100, ROSF consistently outperforms BM25 in terms of both Precision@10 and NDCG@10. ROSF achieves the best performance when the value equals to 70.

### 6.2 Answer to RQ2

RQ2 is examined in our second experiment, which was designed to compare our proposed method ROSF with the comparative methods. Four comparative methods used are: BM25, BM25F, Portfolio [29], and VF [17]. The reason why these methods are selected is that they take free-form queries as inputs and output code snippets for the queries. Specifically, BM25F refers to the strategy that searches multiple fields on Lucene like SSI [7], where BM25 is used to calculate the similarities. BM25F is employed to illustrate the advantage of adjusting automatically for the weights of various features in our method. VF refers to the work of Keivanioo et al. [17], which employs VSM and frequent item-set mining. We also compare our method with Portfolio [29] that combines VSM, PageRank, and SAN.

For drawing confident conclusions whether one algorithm outperforms another, we conduct statistical tests to compare the average results of two metrics for ROSF and the comparative methods. Specifically, first, Friedman's test is employed to detect the potential differences in the performance among the methods. Then, to analyze both the strength and the weakness of the method, we conduct the two-sided Wilcoxon's signed rank tests between ROSF and the other methods. For two statistical tests, when comparing each pair of methods, the primary null hypothesis is that there is no statistical difference in the performance between two methods. In this section, for both Friedman's test and Wilcoxon's test, we adopt the 95% confidence level, *i.e.*, the p−values below 0.05 are considered significant.

Fig. 5 shows the average ranking with respect to the Friedman's test [12] of Precision@10 and NDCG@10. In the subfigures, each column represents the ranking of the corresponding method (higher values indicate better performance). In Fig. 5, we can observe that, for both metrics, ROSF obtains all the best rankings. The Friedman's test detects significant differences in the performance among the methods (with *p-values* = 0.001, and 0.048 for the Precision and NDCG, respectively).

Table 6 presents the results of Wilcoxon's tests between ROSF and the comparative methods. Table 7 shows the details of extremal values, median, mean, and standard deviation of Precision and NDCG. Fig. 6 shows the statistical summary of Precision or NDCG.

For Precision, among the pairwise comparisons in Table 6, we can observe that the *p-values* are all less than 0.05. We reject the null hypothesis and accept the alternative hypothesis that there is a statistically significant difference in the mean value of Precision and NDCG between ROSF and the comparative methods (BM25, Portfolio, VF, or BM25F). In Table 7 and Fig. 6, we can observe that ROSF consistently outperforms the comparative methods in terms of Precision. Specifically, the improvement of ROSF over BM25 is 16%, BM25F is 80%, Portfolio is 20%, and VF is 41%. Considering the family-wise error rate [12], we can deduce that ROSF performs the best among BM25, BM25F, Portfolio, and VF with a p−value less than 1−(1−0.012)×(1−0.0059)×(1−0.0147)×(1−0.0019) < 0.035.

For NDCG, we observe a similar phenomenon except for ROSF vs BM25, where the *p-value* is 0.062. Here, the confidence level is 90%, rather than 95%. In the other pairwise comparisons, ROSF Specifically outperforms the comparative methods with a p−value less than $1 − (1 − 0.0486) \times (1 − 0.0057) \times (1 − 0.0333) < 0.09$. Specifically, the improvement of ROSF over BM25 on NDCG is 12%, BM25F is 24%, Portfolio is 13%, and

TABLE 7
THE STATISTICAL SUMMARY OF THE SECOND EXPERIMENT

| | Approach | Samples | Min | Max | Median | Mean | StdDev |
|---|---|---|---|---|---|---|---|
| Precision@10 | ROSF | 20 | 20% | 100% | 70% | 66.5% | 0.2390 |
| | BM25 | 20 | 0% | 100% | 60% | 57.5% | 0.2552 |
| | Portfolio | 20 | 0% | 90% | 60% | 55.5% | 0.2350 |
| | VF | 20 | 0% | 100% | 50% | 47% | 0.3278 |
| | BM25F | 20 | 0% | 100% | 30% | 37% | 0.2755 |
| NDCG@10 | ROSF | 20 | 0.3082 | 1 | 0.9041 | 0.8448 | 0.1646 |
| | BM25 | 20 | 0 | 1 | 0.7772 | 0.7551 | 0.2407 |
| | Portfolio | 20 | 0 | 1 | 0.7795 | 0.7445 | 0.2325 |
| | VF | 20 | 0 | 0.9816 | 0.8220 | 0.6347 | 0.3526 |
| | BM25F | 20 | 0 | 1 | 0.7316 | 0.6819 | 0.2922 |

TABLE 8
THE VALUES OF SPEARMAN'S RHO BETWEEN NINE FEATURES

|    | f1 | f2    | f3    | f4    | f5    | f6    | f7    | f8    | f9     |
|----|----|-------|-------|-------|-------|-------|-------|-------|--------|
| f1 | 1  | 0.158 | 0.119 | 0.152 | 0.310 | 0.436 | 0.350 | 0.588 | -0.041 |
| f2 |    | 1     | 0.718 | 0.558 | 0.091 | 0.085 | 0.063 | 0.124 | -0.151 |
| f3 |    |       | 1     | 0.492 | 0.156 | 0.017 | 0.044 | 0.118 | -0.132 |
| f4 |    |       |       | 1     | 0.095 | 0.133 | 0.059 | 0.120 | -0.113 |
| f5 |    |       |       |       | 1     | 0.274 | 0.295 | 0.209 | -0.008 |
| f6 |    |       |       |       |       | 1     | 0.157 | 0.402 | 0.029  |
| f7 |    |       |       |       |       |       | 1     | 0.274 | 0.016  |
| f8 |    |       |       |       |       |       |       | 1     | 0.003  |
| f9 |    |       |       |       |       |       |       |       | 1      |

TABLE 9
THE IMPACT OF EACH FEATURE IN THE PERFORMANCE OF OUR METHOD

| Ranking Methods | Precision@10 | | NDCG@10 | |
|---|---|---|---|---|
| | Avg. | Impact | Avg. | Impact |
| ROSF | 66.5% | - | 0.844821 | - |
| ROSF(except f1) | 59.0% | -11.28% | 0.833229 | -1.37% |
| ROSF(except f2) | 66.0% | -0.75% | 0.847242 | +0.29% |
| ROSF(except f3) | 66.0% | -0.75% | 0.848619 | +0.45% |
| ROSF(except f4) | 65.5% | -1.50% | 0.852132 | +0.87% |
| ROSF(except f5) | 66.0% | -0.75% | 0.848438 | +0.43% |
| ROSF(except f6) | 64.0% | -3.76% | 0.868522 | +2.81% |
| ROSF(except f7) | 65.5% | -1.50% | 0.845687 | +0.10% |
| ROSF(except f8) | 64.0% | -3.76% | 0.830116 | -1.74% |
| ROSF(except f9) | 57.0% | -14.29% | 0.757598 | -10.32% |

VF is 33%.

In summary, ROSF improves Portfolio and VF by 20%-41% in Precision@10 and 13%-33% in NDCG@10. The reason may be due to the two stages in ROSF, *i.e.*, coarse-grained searching and fine-grained re-ranking. The first stage focuses on selecting as many as possible relevant code snippets in the candidate set. The second stage tries to rank the relevant code snippets to the top of the recommendation result. Meanwhile, our method also benefits from that the weights of nine features are adjusted automatically by the prediction model.

*Answer to RQ2*: Based on the above observations, we can argue that ROSF is a better method for code snippets recommendation than the comparative methods. This results clearly validate the ability of ROSF for recommending more relevant code snippets in the result, and putting the relevant snippets on higher positions than the irrelevant ones.

### 6.3 Answer to RQ3

Following the previous work [36], we first analyze the correlation between nine features using Spearman's rank correlation coefficient (Spearman's rho) to manage these features. It is appropriate to use Spearman's rho when the relationship between the variables is not linear. High correlation between features makes it difficult to determine the effect of each feature on performance [45]. It is beneficial to minimize the correlated features for speeding up the training process. The values of Spearman's rho range from -1 and +1. Generally, the absolute value greater than 0.6 is considered to be a high level of correlation. If the value of two features is greater than 0.8, it is necessary to remove one of them [9].

Table 8 shows the values of Spearman's rho between nine features of 1050 instances in the training set. This table shows that almost all correlation values between two features are less than 0.6, which means that these features are uncorrelated between them [9]. In this table, we also find that the value between the features f2 and f3 (The textual similarity between a query and the full title and the simple title of a candidate code snippet) is 0.718, which indicates there is a high level of correlation (0.6 - 0.8) between them, but not a very high level (over 0.8) [9]. Moreover, the following experiment also shows that the impacts of the features f2 and f3 on the performance are different. They can't replace each other.

For determining the impact of each feature on the performance of our method, we regard our method with nine features as the baseline. Then, we construct an alternative method for each of nine features by removing that feature. Totally, we build nine alternative methods for nine features. By comparing the performance of the alternative methods against the baseline, we can analyze the impact of each feature on the performance.

Table 9 shows the comparison results between the alternative methods and the baseline in terms of Precision@10 and NDCG@10. We observe that two features, the feature f9 (the number of lines in the code snippets) and the feature f1 (the textual similarity between a query and the content of a candidate code snippet) decrease the performance in terms of Precision@10 significantly. The feature f9 decreases the performance in terms of NDCG@10 significantly. However, other features have slight impacts on the performance. Among them, the features f6 (the textual similarity between a query and the import statements of the Java standard library) and f8 (the topic similarity) have more impacts than others.

*Answer to RQ3*: The impact on the performance is different for nine features. The following features are influential on the performance of ROSF: f9, f1, f6, f8, *i.e.*, the number of lines (f9), the textual similarity between queries and contents (f1), the textural similarity between a query and the import statement of Java library (f6), and the topic similarity between a query and the content of a candidate code snippet (f8).

## 7 THREATS TO VALIDITY

This section discusses threats to validity of our work.

*The query set:* The first major factor that influences the performance is the query set. Different methods may have different performance for the same query. In order to reduce bias in comparing these different methods, 20 same queries are used in our experiments.

This size is similar to previous work [17], [7], [10]. However, this may still threaten the validity of the results. Moreover, the performance of a method also relies on how good the keywords in the queries are. It is also our further work for choosing adequate words as queries.

*Code snippet corpus:* Our several experiments are all conducted on a real-world code snippet corpus of open source mobile app projects. Although this corpus has a certain scale, in comparison to the millions of apps in Google play, it is relatively small. Meanwhile, the features used in our method can also be identified in other desktop software. But we have not tested it by ourselves. We plan to evaluate the effectiveness of our method on different size of corpora for different Internetware software. Moreover, in our future work, we also plan to recommend different granularities of code artifacts (*e.g.*, Java files) for the same query.

*Comparative methods:* In the second experiment, we reproduce several previous methods for comparison. There are certain gaps in the performance between the reproduced methods and the original ones. The possible reason may be the difference in the code snippet corpora. The corpora in previous studies were collected from desktop software, and our corpus is collected from Android app projects. The methods may be unfit for our code snippet corpus. For example, the traditional static control-flow analysis cannot be directly applied to Android apps, because the apps are framework-based and event-driven [43]. Moreover, the effectiveness of the call graph [43] and the usage similarity [7] for our method will be explored in our next work.

## 8. RELATED WORK

In this section, we show the related studies with our work. In recent years, several studies are presented to support the automatic recommendation of code examples for different types of inputs. Table 10 provides the comparisons between ROSF and other methods from several different angles.

*Free-form query:* Similar to the following studies, our work also uses the free-form queries as inputs. However, these studies are mainly based on information retrieval techniques, while our method combines both information retrieval and supervised learning to recommend code snippets. For example, Bajracharya et al. [7] propose a structural Semantic Indexing (*SSI*) to recommend source code entities (classes, methods, etc.) based on the similarities of APIs usage. This technique is implemented on Lucene, where the boost values of index fields need to be set manually before searching. In contrast to our method, the weights of several features employed in our experiments are adjusted automatically by the prediction model. McMillan et al. [29] propose a code search system called *Portfolio*, which can find relevant functions that implement the given queries, and show the visualizing dependencies of the retrieved functions. This system combines NLP, PageRank, and SAN algorithms. Keivanloo et al. [17] present a method for spotting working code examples by combining p-strings and VSM with frequent item-set mining. Lv et al. [23] propose CodeHow, a code search technique by considering both API understanding and textual similarity matching. The evaluation results based on C# projects show that CodeHow achieves a precision score of 0.794. In this method, the online documents of APIs are used to expand query. However, these

TABLE 10
COMPARISON OF ROSF WITH OTHER RELATED METHODS

| Approach | Year | Input Type | Output Type | Information | Search Method | Tool | Reference |
|---|---|---|---|---|---|---|---|
| SSI | 2010 | FQ | C | FCC, FQN, T, TU, J | WM, MF | L | [7] |
| Portfolio | 2013 | FQ | C, CC | FCC, CG | PR, SA, WM | L | [29] |
| VF | 2014 | FQ | C | FCC | WM, FIM | - | [17] |
| MAPO | 2009 | N | C, UP | CS | WM, FIM | - | [54] |
| Baker | 2014 | N | C | SC | DL | iAST | [46] |
| MUSE | 2015 | N | C | SE,CS | SS, CD, H | - | [33] |
| PARSEWeb | 2007 | OT | C | FCC, CS | WM, Q | AST | [27] |
| Strathcona | 2005 | C | C | SC | H | - | [14] |
| Xsnippet | 2006 | C | C | SC, CG | T, W, PT | - | [42] |
| Ichi Tracker | 2012 | C | C | FCC, CH | CD, CHT | S, G, K, CCF | [15] |
| ROSF | - | FQ | C | FCC, FQN, T, NSM, SC | WM, TM, LR | L, W | - |

*Column Input specifies the input type for each method (Free-form query (FQ), API name (N), Code Snippet (C), or the object type of source and Destination (OT)). Column Output specifies the output type for each method (Code Snippet (C), call chains (CC), or API Usage Patterns (UP)). Column Information specifies the information used for each method (Full text of the Code and Comments (FCC), Fully Qualified Name (FQN), Javadoc (J), Title of entities (T), Title of other entities that have similar Usage of API (TU), Names of Sibling Methods (NSM), Call Graph (CG), Structural Characteristics (SC), API method call sequences (CS), similar examples (SE), or Code History (CH)). Column Search Method specifies the methods used for each method (PageRank (PR), Spreading Activation Network (SA), Word Matching (WM), Parameter Type matching (PT), Query expansion techniques (Q), Multiple Fields searching (MF), Frequent Item-set Mining (FIM), Heuristics(H), Deductive Linking(DL), Static Slicing(SS), Clone Detection (CD), Code History Tracking(CHT), Topic Model(TM), Logistic Regression(LR)). Column Tool specifies the tool or platform used for each method (Lucene (L), incomplete Abstract Syntax Tree (iAST), Abstract Syntax Tree (AST), SPARS/R (S), Google Code Search (G), Koders (K), CCFinder (CCF), or WEKA (W)).*

documents are not always available [16].

*API name:* Moreover, several studies focus on solving API usage problem [46]. They take API method names as inputs, and output code examples [36], concise method usages [33], or API usage patterns for special API methods [26], [54]. For example, In 2009, Zhong et al. [54] present an API usage mining framework called *MAPO*, which mines and recommends API usage patterns and code snippets for given requests from developers. In 2014, Subramanian et al. [46] propose a method, *Baker*, to enhance traditional API documentations with up-to-date code examples. The purpose of recommendation is to provide reusable code snippets for developers. For showing how to use a specific method, Moreno et al. [33] present a method named *MUSE* by combining static slicing and clone detection technology to provide concise examples for that method. Each example contain the sequence of relevant steps to invoke the method, and the less relevant code is pruned out. In 2016, Niu et al. [36] propose a code example search approach applying a machine learning technique (*i.e.*, learning to rank) to recommend code examples taking method names and class names as inputs. Different from these studies, our method takes free-form queries as inputs and outputs the original code snippets to developers.

*Others:* Moreover, some studies employ other forms of queries as input to recommend code examples, such as the pair of types [27], the code samples [42], and test cases [19]. For example, Mandelin et al. [27] provide a code search engine called *PARSEWeb* using the query in the form "*Source-Destination*". Holmes et al. [14] propose *Strathcona* Example Recommendation Tool to assist developers. *Strathcona* can employ structural characteristics of both the past projects and the developers' current context to automatically recommend relevant examples. To decrease the number of irrelevant results, *Xsnippet* [42] improve *Strathcona* by employing the graph mining technique. Meanwhile, a combination of popularity, size, and context is employed to improve the ranking. In 2012, Inoue et al. [15] propose a prototype named *Ichi Tracker*, which takes a code fragment as its input, and returns the code fragments. Except for the studies recommending code examples, there also exists some work to recommend the method call sequences and the relevant APIs, such as *Sourcerer* [20], and *Export* [49].

Except the above studies for recommending code snippets and APIs, our work is also related with automatic patch generation [18], [48] and automated program repair [39]. These studies all generate promising results. For example, Kim et al. [18], propose PAR, an automated patch generation technique, by leveraging the fix patterns learned from existing human-written patches. Tao et al. [48], conduct a large-scale human study to investigate the usefulness of automatically generated patches as debugging aids. Different from the studies above, which automatically generate patches or repair bugs according to code context or bugs, our method only provides relevant code snippets to developers for reuse according to the free-form queries.

## 9. CONCLUSION AND FUTURE WORK

In this paper, we propose a method called ROSF based on information retrieval and supervised learning to recommend relevant code snippets for the given free-form queries. We identify nine features to generate the instances in the training set and the test set. To evaluate the effectiveness of our method, several experiments are conducted on a real-world code snippet corpus. These code snippets come from 1,538 open source app projects. The results of these experiments state that our method is effective for code search, and outperforms the previous state-of-the-art methods by 20%-41% in Precision@10 and 13%-33% in NDCG@10.

We consider two aspects as our future work. The first is providing more resources for app development. In addition to code snippets, other entities (*e.g.*, permissions, screenshots) and relations (*e.g.*, call graph, API usages) are also important for implementing the programming tasks. The second is exploiting more domain features. Our method employs nine features to characterize code snippets. These features can also be extracted in desktop software. In the future work, for improving the performance of recommendation, we plan to identify other special information in the field of mobile apps, such as user reviews and the descriptions of apps, and other features of Internetware software.

## ACKNOWLEDGMENTS

This work is supported in part by the National Program on Key Basic Research Project under Grant 2013CB035906, the New Century Excellent Talents in University under Grant NCET-13-0073, the National Natural Science Foundation of China under Grants 61370144, 61403057, and 61572097, the Fundamental Research Funds for the Central Universities under Grant DUT14YQ203 and DUT14RC(3)150.

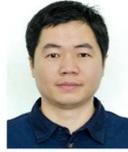
**He Jiang**, received the Ph.D. degree in computer science from the University of Science and Technology of China, Hefei, China. He is currently a Professor with the Dalian University of Technology, Dalian, China. His current research interests include search based software engineering and mining software repositories. Dr. Jiang is also a member of the ACM and the CCF.

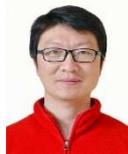
**Liming Nie**, received the M.Sc. degree in Guangxi University for Nationalities, Nanning, China, in 2009. He is currently a Ph.D. candidate in Dalian University of Technology. His current research interests include code recommendation and data mining in software engineering.

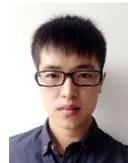
**Zeyi Sun**, received the B.Sc. degree in software engineering from the Dalian University of Technology, Dalian, China, in 2015. He is currently a Master of Software Engineering candidate in Dalian University of Technology. His current research interest is code recommendation in software engineering.

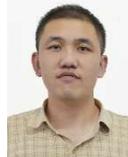
**Zhilei Ren**, received the B.Sc. degree in software engineering and the Ph.D. degree in computational mathematics from the Dalian University of Technology, Dalian, China, 2013, respectively. He is currently a lecturer with the Dalian University of Technology. His current research interests include evolutionary computation and its applications in software engineering. Dr. Ren is a member of the ACM and the CCF.

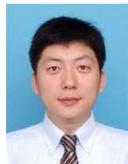
**Weiqiang Kong**, received the Ph.D. degree in information science from Japan Advanced Institute of Science and Technology, Japan. He is currently a professor with the Dalian University of Technology, Dalian, China. His research interests include formal methods, in particular, formal verification with hybrid model checking techniques for software analysis.

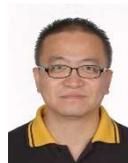
**Tao Zhang**, received the Ph.D. degree in Computer Science from University of Seoul, South Korea in Feb., 2013. He was a postdoctoral fellow at the Department of Computing, Hong Kong Polytechnic University from November 2014 to November 2015. Currently, he is an assistant professor at the School of Computer Science & Technology, Nanjing University of Posts and Telecommunications. His research interest includes data mining, software maintenance and natural language processing.

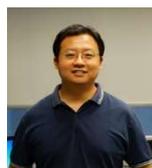
**Xiapu Luo**, received his Ph.D. degree in Computer Science from the Hong Kong Polytechnic University. After that, Dr. Luo spent two years at the Georgia Institute of Technology as a post-doctoral research fellow. Currently, he is a research assistant professor at the Department of Computing, Hong Kong Polytechnic University. His research interests include Software Analysis, Android Security and Privacy, Cloud Computing, and Mobile Networks.